\documentclass[traditabstract]{aa} 
\usepackage{graphicx}
\usepackage{natbib}
\usepackage{txfonts}

\begin{document}

\title{The cool magnetic DAZ white dwarf NLTT~10480\thanks{Based on
observations collected at the European Organisation for Astronomical Research
in the Southern Hemisphere, Chile under programme ID 080.D-0521, 082.D-0750 and 
086.D-0562.}}

\author{Ad\'ela Kawka \and St\'ephane Vennes}

\institute{Astronomick\'y \'ustav, Akademie v\v{e}d \v{C}esk\'e republiky, Fri\v{c}ova 298, CZ-251 65 Ond\v{r}ejov, Czech Republic\\
\email{kawka,vennes@sunstel.asu.cas.cz}
}

\date{Received; accepted}

\abstract{We have identified a new cool magnetic white dwarf in the New Luyten 
Two-Tenths (NLTT) catalogue. The high proper-motion star NLTT~10480 
($\mu=0.5\arcsec$\,yr$^{-1}$) shows weak Zeeman-split lines of calcium as well 
as characteristic H$\alpha$ and $\beta$ Zeeman triplets. Using VLT X-shooter 
spectra, we measured a surface-averaged magnetic field $B_S\sim 0.5$~MG. 
The relative intensity of the $\pi$ and $\sigma$ components of the calcium and 
hydrogen lines imply a high inclination ($i\ga 60^\circ$). The 
optical-to-infrared $V-J$ colour index and the \ion{Ca}{i}$/$\ion{Ca}{ii} 
ionization balance indicate a temperature between 4\,900 and 5\,200\,K, while 
the Balmer line profiles favour a higher temperature of 5400\,K. The 
discrepancy is potentially resolved by increasing the metallicity to 
$0.03\times$ solar, hence increasing the electron pressure. However, the 
measured calcium abundance and abundance upper limits for other elements 
(Na, Al, Si. and Fe) imply a low photospheric metallicity $\la 10^{-4}$ solar.  
Assuming diffusion steady-state, a calcium accretion rate of 
$\log{\dot{M}({\rm g\,s}^{-1})}=5.6\pm0.3$ is required to sustain a calcium
abundance of $\log{n{\rm (Ca)}/n{\rm (H)}}=-10.30\pm0.30$ in the white dwarf 
atmosphere. We examine the implications of this discovery for the incidence of 
planetary debris and weak magnetic fields in cool white dwarf stars.}

\keywords{white dwarfs -- stars: individual: NLTT~10480 -- stars: atmospheres 
-- stars: abundances}

\maketitle

\section{Introduction}

NLTT~10480 (LHS~5070, LP~887-66) is a high proper-motion star 
\citep{luy1979,luy1980} that was also listed as a white dwarf candidate in 
Luyten's white dwarf catalogue \citep{luy1977}.
Using an optical-infrared reduced proper-motion diagram \citep{sal2002} and
additional colourimetric criteria \citep{kaw2004b}, we selected NLTT~10480 for
spectroscopic observations to investigate stellar properties such as 
the effective temperature ($T_{\rm eff}$), surface gravity ($\log{g}$),
chemical composition, and magnetic field strength, and to constrain its cooling 
age and mass. 

The chemical composition of cool white dwarfs shows great diversity. Heavy 
elements, particularly calcium, are detected in close to a quarter of cool, 
hydrogen-rich (DA) white dwarfs, but with abundances well below solar 
\citep{zuc2003}. The abundance of heavy elements decreases with white dwarf 
cooling ages mainly because of the increasing depth of the mixed convective 
layers in aging white dwarfs \citep{paq1986,koe2009}. Few examples of very cool 
($T_{\rm eff}\la 5\,000$ K) polluted white dwarfs (DAZ) are known, such as
G~77-50 \citep[WD~0322$-$019,][]{hin1974,sio1990}, which is also harbouring a 
weak magnetic field \citep{far2011}. New high-dispersion and high 
signal-to-noise ratio spectroscopic observations of faint high proper-motion 
stars are likely to contribute new objects to the current sample.

Heavy elements in the atmosphere of cool white dwarfs are almost certainly 
accreted from their immediate environment. \citet{kil2006} and \citet{far2009} 
reported infrared observations of a sample of cool white dwarfs contaminated 
with heavy elements, and the authors noted an infrared-excess incidence of 
$\approx$10-20\%. 
This excess was attributed to debris discs of temperatures ranging from a few 
hundred degrees to over 1000~K. Some cool DAZ white dwarfs, such as G~174$-$74 
(WD~0245+541), do not show an infrared excess \citep{deb2007}, and the presence 
of heavy elements in cool white dwarfs with ages in excess of several billion 
years suggests, instead, the effect of episodic accretion from small asteroids 
rather than from a stable debris disc \citep{jur2008}. Therefore, the 
identification of new cool DAZ white dwarfs is of interest to constrain the 
phenomenon.

We present a first report on a programme aimed at identifying and characterizing
new DAZ white dwarfs in the New Luyten Two-Tenths (NLTT) catalogue. Sect. 2
describes observations obtained at the European Southern Observatories (ESO)
using the New Technology Telescope (3.6-m) and the Very Large Telescopes (VLTs).
Sect. 3 presents our model atmosphere analysis including details of the model 
structures (Sect. 3.1), heavy element line opacities (Sect. 3.2), and Zeeman 
effect on line profiles (Sect. 3.3). We summarize and discuss some implications 
of our results in Sect. 4.

\section{Observations}

\begin{table}
\caption{Log of spectroscopic observations\label{tbl-log}}
\centering
\begin{tabular}{llll}
\hline\hline
Instrument & UT Date & Range & Note\tablefootmark{1} \\
\hline\\
VLT/FORS1     & 2007 Nov 1   & 3780-6180\AA       & $\Delta\lambda\sim$6\AA  \\
NTT/EFOSC2    & 2008 Oct 23  & 3680-7400\AA       & $\Delta\lambda\sim$14\AA \\
VLT/X-shooter & 2010 Dec 10  & 3000\AA-2.5$\mu$m  &  $R\sim9000$ \\
              & 2011 Jan 10  &                    &              \\
              & 2011 Jan 26  &                    &              \\
              & 2011 Mar 3   &                    &              \\
\hline
\end{tabular}
\tablefoottext{1}{$\Delta\lambda\equiv {\rm FWHM}$, $R\equiv \lambda/\Delta\lambda$.}
\end{table}

We first observed NLTT~10480 with the focal reducer and low-dispersion
spectrograph (FORS1) attached to the 8m UT2 (Kueyen) at Paranal Observatory
as part of our spectro-polarimetric survey of white dwarfs. The purpose
of the survey was to search for white dwarfs with weak magnetic fields.
We used the 600B grism combined with a slit-width of 1 arcsecond that provided a 
resolution of 6.0 \AA. The spectra covered the range between 3780 and 6180 \AA.
The observations were conducted on UT 2007 Nov 1 and consisted of a 
sequence of two consecutive exposures with an exposure time of 1360 s each. 
In the first exposure the Wollaston prism is rotated to $-45^\circ$ and it is followed by a 
second exposure with the Wollaston prism rotated to $+45^\circ$ from which we
extracted the flux and circular polarization spectra.

Since the FORS1 spectra did not include H$\alpha$, we obtained two additional 
low-dispersion spectra with the ESO Faint Object Spectrograph and Camera
(EFOSC2) attached to the New Technology Telescope (NTT) at La Silla Observatory 
on UT 2008 Oct 23. We used Grism 11 which has 300 lines per mm and a blaze 
wavelength of 4000 \AA. The slit-width was set to 1 arcsecond, which resulted 
in a spectral resolution of $\sim 14$ \AA. The exposure time of each spectrum 
was 1500 seconds. Observations were carried out at the parallactic angle and
were flux calibrated with the flux standard Feige 110.

\begin{figure}
\includegraphics[width=1.00\columnwidth]{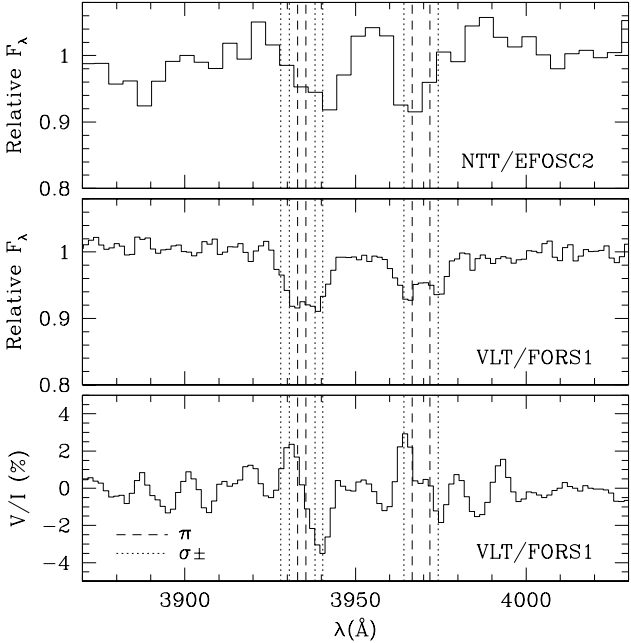}
\caption{Flux spectra of the \ion{Ca}{II} H\&K lines in NLTT~10480 obtained with EFOSC2 ({\it top}) and FORS1 ({\it middle}),
and circular polarization ({\it bottom}) spectrum obtained with FORS1.
The dotted lines mark the position of the $\sigma$ 
components and the dashed lines mark the $\pi$ components (see Sect. 3.2.3).
\label{fig-spol}}
\end{figure}

Figure~\ref{fig-spol} compares the low-resolution spectrum of \ion{Ca}{II} lines
obtained with EFOSC2 to the higher resolution spectrum of FORS1. The flux and 
circular polarization spectra clearly show the $\sigma$ components, proving 
NLTT~10480 to be magnetic.

Finally, we obtained a set of  echelle spectra of NLTT~10480 with the
X-shooter spectrograph attached to the UT2 (Kueyen) at Paranal Observatory.
The spectra were obtained at four epochs (UT 2010 Dec 10, 2011 Jan 9 and 26, 
and March 3). The 2011 Mar 3 spectrum was of much poorer quality and was 
excluded from the analysis. The spectra were obtained with the slit-width set 
to 0.5, 0.9, and 0.6 arcsecond for the UVB, VIS, and NIR arms, respectively. This 
set-up delivered a resolving power of 9100 for UVB, 8800 for VIS, and 6200 for 
NIR. The exposure time for each exposure was 2400 seconds.

Table~\ref{tbl-log} summarizes our spectroscopic observations.
The signal-to-noise ratio in the summed X-shooter spectrum is $S/N\approx53$ 
per binned-pixel at 4000\AA, while it is $\approx 120$ in the summed FORS1 
spectrum and $\approx 65$ in the summed EFOSC2 spectrum at the same wavelength.

We searched for photometric measurements using VizieR. The Two Micron All Sky 
Survey (2MASS) listed infrared $JHK$ magnitudes, but only the $J$ magnitude
was of acceptable quality. We also used the acquisition images from the 
X-shooter observations to estimate a $V$ magnitude for NLTT~10480. We used 11 
acquisition images of Feige 110 obtained between UT 2010 Dec 10 and 2011 Jan 1, 
and set the zero point. Next, we determined an average $V$ magnitude for 
NLTT~10480 using our five acquisition images. We employed the atmospheric 
extinction table of \citet{pat2011}. Table~\ref{tbl-phot} lists the photometric 
measurements.

\begin{table}[t!]
\caption{Photometry and astrometry\label{tbl-phot}}
\centering
\begin{tabular}{lcc}
\hline\hline
Band & Measurement & Reference \\
\hline
$V$   & $17.49\pm0.05$   & 1 \\
$J$   & $16.003\pm0.078$ & 2 \\
$H$   & $15.836\pm0.182$ & 2 \\
$K_S$ & $16.529$:        & 2 \\
      &                  &   \\
RA (J2000.0)  &  03 17 12.08 & 3 \\
Dec (J2000.0) & $-$29 11 34.33 & 3 \\
($\mu$,\ $\theta$) & ($0.496\pm0.008$ mas\,yr$^{-1}$,\ 162$^\circ$) & 3 \\
                   & ($0.503\pm0.007$ mas\,yr$^{-1}$,\ 159$^\circ$) & 4 \\
                   & ($0.49 \pm0.01 $ mas\,yr$^{-1}$,\ 160$^\circ$) & 5 \\
\hline
 & & \\
\end{tabular}
\tablebib{(1) Johnson $V$, this work; (2) 2MASS \citep{skr2006};
(3) Revised NLTT \citep{sal2003}; (4) SPM Catalog 2.0 \citep{pla1998}; (5) Liverpool-Edinburgh High Proper Motion Catalogue \citep{pok2004}.}
\end{table}

The SPM Catalog provides a photographic $B$ magnitude ($18.08\pm0.16$). 
However, due to large uncertainties in the SPM and other photographic 
magnitudes, these are not as useful as our $V$ magnitude and the 2MASS $J$ 
magnitudes. The 2MASS $K$ magnitude is unreliable and the $H$ magnitude is 
uncertain. The index $B-V=0.59\pm0.17$ loosely constrains the temperature in 
the range $5\,100\la T_{\rm eff}\la 6\,700$ K, but the more precise index 
$V-J=1.49\pm0.09$ implies that NLTT~10480 is a cool white dwarf with 
$T_{\rm eff}\approx 5\,050\pm150$ K (see Table~\ref{tbl-model} and Sect. 3.1).

NLTT~10480 is also characterized by a large proper-motion of 
$\sim0.5\arcsec$\,yr$^{-1}$ (Table~\ref{tbl-phot}). The kinematical properties 
were determined using the proper-motion and the X-shooter radial velocity 
measurement (see Sect. 4).

\section{Analysis}

\subsection{Model atmospheres and spectral syntheses}

We calculated a grid of model atmospheres for cool hydrogen-rich white dwarfs.
The models are in convective and radiative equilibrium with the total flux
converged to better than 0.01\% in all layers. The model grid covers
the effective temperature range $4\,900\le T_{\rm eff}\le 6\,000$ K in
100 K steps 
and the surface gravity range  $7.5\le \log{g}\le8.5$ in steps of 0.25 dex.
The adopted treatment of the convective energy transport and pressure ionization effects are
described in \citet{kaw2006}, but further improvements to the models will be described in
a forthcoming publication.
All relevant species (H, H$^+$,  H$_2$,  H$_2^+$,  H$_3^+$) are included in the statistical 
equilibrium equation. We employed the H$_3^+$ partition function of \citet{nea1995}. 
Electrons contributed by identifiable trace elements (e.g., calcium) are also
included in the charge conservation equation, although the ionization of hydrogen atoms 
and molecules dominate the electron budget.  

The model atmospheres include opacities caused by H bound-bound, 
bound-free and free-free transitions, H$^-$ bound-free and free-free 
transitions, H$_2-$H$_2$ collision-induced absorption 
\citep[CIA,][]{bor1997,bor2001}, and the H$_2$-H and H-H
collision-induced absorptions in the far Ly$\alpha$ wing 
\citep[see][]{kow2006} using opacity tables from \citet{roh2011}.
Finally, the H$_2$ and H Rayleigh scattering are included along with electron scattering.

Synthetic colours as well as detailed hydrogen and heavy element line profiles 
are computed using the model structures. Table~\ref{tbl-model} lists some 
photometric properties of the cool models. The colour indices at shorter 
wavelengths are effected by the Ly$\alpha$ collision-induced absorptions.
\citet{all2008} calculated new H$\alpha$ line profiles including 
self-broadening effects and found a half-width at half maximum (HWHM)
$\approx40$\% larger than estimated by \citet{ali1965,ali1966} and comparable
to the calculations of \citet{bar2000a}. We adopted the H$\alpha$ HWHM from
\citet{all2008} and the H$\beta$, and H$\gamma$ cross-sections and velocity
parameters, converted into HWHM, from \citet{bar2000a}. We describe the heavy 
element line profiles in the following section.

\begin{table}[t!]
\caption{Selected synthetic colours \label{tbl-model}}
\centering
\begin{tabular}{lccc}
\hline\hline
$\log{g}$ & $T_{\rm eff}$ & $B-V$ & $V-J$ \\
  (cgs)   &      (K)    & (mag) & (mag) \\
\hline
7.5      & 4\,900        & 0.826 & 1.570 \\
         & 5\,100        & 0.733 & 1.460 \\
         & 5\,300        & 0.658 & 1.360 \\
	 & 5\,500        & 0.602 & 1.270 \\
	 & 5\,700        & 0.556 & 1.188 \\
8.0      & 4\,900        & 0.858 & 1.562 \\
         & 5\,100        & 0.760 & 1.453 \\
         & 5\,300        & 0.677 & 1.351 \\
	 & 5\,500        & 0.611 & 1.260 \\
	 & 5\,700        & 0.559 & 1.178 \\
8.5      & 4\,900        & 0.887 & 1.550 \\
         & 5\,100        & 0.790 & 1.446 \\
         & 5\,300        & 0.701 & 1.344 \\
	 & 5\,500        & 0.627 & 1.251 \\
	 & 5\,700        & 0.568 & 1.167 \\
\hline
\end{tabular}
\end{table}

\subsection{Neutral and ionized line profiles}

\subsubsection{Line broadening}

The dominant broadening mechanism is collision with hydrogen atoms. We employed
the coefficients of \citet{bar2000b}, where the full-width at half-maximum (FWHM) of the
Lorentzian profiles is given by
\begin{equation}
\frac{w}{n(\ion{H}{i})} = \Big{(}\frac{T}{10^4{\rm K}}\Big{)}^{(1-\alpha)/2}\, \Gamma\ \ {\rm rad\,s^{-1}\,cm^3},
\end{equation}
where $\log{\Gamma} = -7.562$ for \ion{Ca}{i}$\lambda4226$ and $-7.76$ for \ion{Ca}{ii} H\&K at
$T=10\,000$ K, and $\alpha = 0.238$ for \ion{Ca}{i}$\lambda4226$ and $0.223$ for \ion{Ca}{ii} H\&K.
Although the adopted broadening parameters do not include the effect of
hydrogen molecules, these provide $>$50\% of the gas pressure in some layers. 
Following the approximate treatment of \citet{kur1981}, the broadening 
parameter $\Gamma\propto n_{\rm H}+0.85\,n({\rm H}_2)$, and, therefore, 
hydrogen molecules may contribute to the total line width. We found that 
neither \ion{Ca}{i} or \ion{Ca}{ii} equivalent widths are significantly 
affected in models at $T_{\rm eff}=5\,400$\,K, but abundances inferred from 
\ion{Ca}{i}$\lambda4226$ may be underestimated by a factor of $\sim 2$ in 
models at 4\,900\,K.

\subsubsection{Zeeman effect}

Fundamentals of stellar line formation in the presence of a magnetic field
are described by \citet{unn1956} and \citet{mar1981}, who also describe 
significant magneto-optical effects. In particular, \citet{unn1956} showed that
the effect of field inclination with respect to to the line-of-sight on the
relative intensity of $\sigma$ and $\pi$ components reaches a maximum at an
angle of 55$^\circ$. Moreover, \citet{mar1981} showed that taking into account
magneto-optical effects may enhance the depth of the $\pi$ components upon 
certain conditions.

\citet{kem1975} studied the quadratic Zeeman effect for the \ion{Ca}{II} H and
K lines and showed that the linear Zeeman effect at fields of 15~MG is still 
dominant. 

The \ion{Ca}{II} H and K lines are the result of transitions between the
ground state with $J,L,S = \frac{1}{2},0,\frac{1}{2}$, and the excited states 
with $J,L,S = \frac{1}{2},1,\frac{1}{2}$ and $\frac{3}{2},1,\frac{1}{2}$, 
respectively, where $J$ is the total angular momentum, $L$ is the orbital 
angular momentum and $S$ is the spin angular momentum. 
The \ion{Ca}{I}$\lambda4226$ line is the result of 
transitions between the ground state with $J,L,S = 0,0,0$
and the excited state with $J,L,S = 1,1,0$.

The levels are split by a magnetic field into $2J+1$ components
defined by the magnetic quantum number $m=-J,...,J$:
\begin{equation}
\Delta \lambda = \frac{eB\lambda^2}{4\pi m_e c^2}(g_l m_l - g_u m_u) \approx 4.67\times10^{-7} \lambda^2 B (g_l m_l - g_u m_u),
\end{equation}
where $\lambda$ is the wavelength in \AA, $B$ is the magnetic field in MG, $e$ 
is the electron charge, $m_e$ is the electron rest mass and $c$ is the speed of
light. The Land\'e factor and the magnetic quantum number of the upper and
lower levels are given by $g_u, m_u$ and $g_l, m_l$, respectively. Land\'e
factors for calcium and other elements except iron were calculated assuming LS 
coupling:
\begin{equation}
g = 1+\frac{J(J+1)-L(L-1)+S(S+1)}{2J(J+1)}.
\end{equation}
For the 4S$_{1/2}$ state of \ion{Ca}{II}, the experimentally determined $g$ 
factor of 2.00225664 \citep{tom2003} agrees with the theoretically 
calculated factor based on the LS coupling scheme. Similarly, for the 
3p\ $^2$P$_{3/2}$ state of \ion{Al}{I}, the experimentally determined $g$ 
factor of $1.33474\pm0.00005$ \citep{mar1968} agrees with the LS 
coupling scheme calculated factor. For iron, the Land\'e factors were obtained
from the Vienna Atomic Line Database (VALD)\footnote{http://vald.astro.univie.ac.at \citep{kup2000}.}.

The permitted transitions are defined by $\Delta m = 0,\pm1$, where 
$\Delta m = 0$ defines the $\pi$ components and $\Delta m = \pm1$ the $\sigma$ 
components.

\subsubsection{Opacity and line intensities}

The relative intensities of the Zeeman components are computed following 
\citet{con1963}. They found for the transitions where $\Delta J = 0$ and 
$\Delta m = 0$ ($\pi$ components)
\begin{equation}
I \propto m^2, 
\end{equation}
or where $\Delta m = \pm1$ ($\sigma$ components)
\begin{equation}
I \propto \frac{1}{4}(J\mp m)(J\mp m+1).
\end{equation}
Similarly, they found for transitions where $J \rightarrow J+1$ and for the 
$\pi$ components
\begin{equation}
I \propto (J+1)^2-m^2,
\end{equation}
and for the $\sigma$ components
\begin{equation}
I \propto \frac{1}{4}(J\pm m+1)(J\pm m+2).
\end{equation}
Finally, for transitions where $J \rightarrow J-1$, the relative intensities 
are for the $\pi$ components
\begin{equation}
I \propto J^2-m^2,
\end{equation}
and for the $\sigma$ components
\begin{equation}
I \propto \frac{1}{4}(J\mp m)(J\mp m-1).
\end{equation}
Table~\ref{tbl-split} lists the total angular momenta, Land\'e factors, 
magnetic quantum numbers for the various transitions of calcium in a magnetic 
field of 0.513~MG. The relative intensities for the different components are 
also provided. 

\begin{figure}[t!]
\includegraphics[width=1.00\columnwidth]{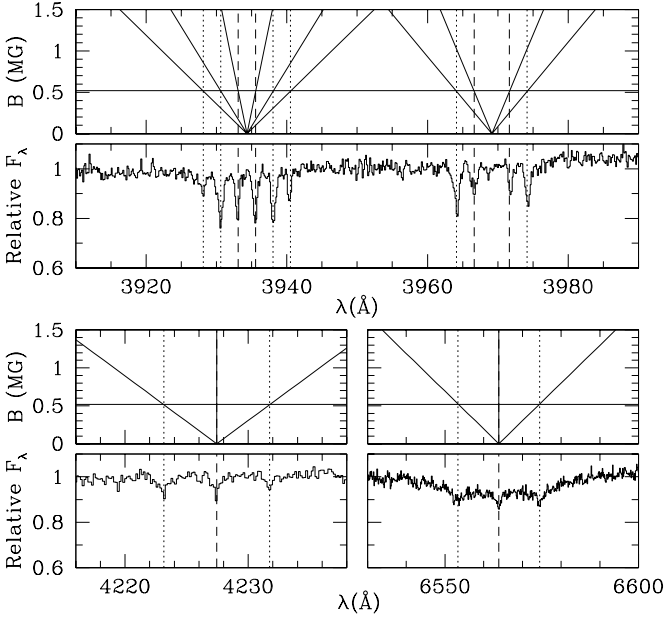}
\caption{Linear Zeeman splitting of \ion{Ca}{II} (H and K) ({\it top}), 
\ion{Ca}{I} 4226 \AA\ ({\it bottom left}), and H$\alpha$ ({\it bottom right}) 
lines at $B=0.519$~MG compared to the observed Zeeman splitting in the spectrum 
of NLTT~10480. The dotted lines indicate the $\sigma$ components and dashed 
lines indicate the $\pi$ components.
\label{fig-Ca_H}}
\end{figure}

\begin{table}[t!]
\caption{Zeeman splitting at $B=0.513$~MG (\ion{Ca}{i} and \ion{Ca}{ii}).\label{tbl-split}}
\centering
\begin{tabular}{ccrcccrcc}
\hline\hline
 \multicolumn{3}{c}{Lower level} & & \multicolumn{3}{c}{Upper level} & $\lambda$ & Rel. Int. \\
 \cline{1-3} \cline{5-7} \\
       $J$    & $g$ & $m$        & & $J$           & $g$         & $m$            & (\AA) & \\
\hline
                      \multicolumn{9}{c}{\ion{Ca}{ii}$\lambda$3933.663} \\
$1/2$         & 2 & $-1/2$       & & $3/2$         & $4/3$         & $1/2$          & 3927.482 & $1/16$ \\
              &   & $1/2$        & &               &               & $3/2$          & 3929.955 & $3/16$ \\
              &   & $-1/2$       & &               &               & $-1/2$         & 3932.427 & $1/4$  \\
              &   & $1/2$        & &               &               & $1/2$          & 3934.899 & $1/4$  \\
              &   & $-1/2$       & &               &               & $-3/2$         & 3937.371 & $3/16$ \\
              &   & $1/2$        & &               &               & $-1/2$         & 3939.844 & $1/16$ \\
\hline
           \multicolumn{9}{c}{\ion{Ca}{ii}$\lambda$3968.469} \\
$1/2$         & 2 & $-1/2$       & & $1/2$         & $2/3$         & $1/2$          & 3963.436 & $1/4$  \\
              &   & $-1/2$       & &               &               & $-1/2$         & 3965.953 & $1/4$  \\
              &   & $1/2$        & &               &               & $1/2$          & 3970.985 & $1/4$  \\
              &   & $1/2$        & &               &               & $-1/2$         & 3973.502 & $1/4$  \\
\hline
           \multicolumn{9}{c}{\ion{Ca}{i}$\lambda$4226.728} \\
$0$           & 0 & $0$          & & $1$           & $1$           & $1$            & 4222.446 & $1/4$  \\
              &   & $0$          & &               &               & $0$            & 4226.728 & $1/2$  \\
              &   & $0$          & &               &               & $-1$           & 4231.010 & $1/4$  \\
\hline
\end{tabular}
\end{table}

The observed relative intensities of the $\pi$ 
and $\sigma$ components will also vary as a function of the angle between the 
magnetic field lines and the line of sight \citep{unn1956,mar1981}.
The monochromatic opacity of the $\pi$ components depends on the angle $\psi$
between the magnetic field axis and the line of sight:
\begin{equation}
\frac{\chi_\pi (\psi)}{\chi_\pi (90^\circ)}\ = \sin^2\psi.
\end{equation}
Similarly, the monochromatic opacity of the $\sigma$ components vary as
\begin{equation}
\frac{\chi_\sigma (\psi)}{\chi_\sigma (90^\circ)}\ = (1+\cos^2\psi),
\end{equation}
where the peak opacities $\chi_\pi (90^\circ)$ and $\chi_\sigma (90^\circ)$ are given at $\psi=90^\circ$ and, as
shown earlier, are calculated following \citet{con1963}.
For example, for hydrogen Balmer lines or \ion{Ca}{i}$\lambda$4226 we have that 
\begin{equation}
\chi_\pi (90^\circ) = \frac{1}{2} \chi_0,\ {\rm and}\ \chi_{\sigma} (90^\circ) = \frac{1}{4} \chi_0.
\end{equation}

Figure~\ref{fig-Ca_H} shows the observed linear Zeeman splitting of 
\ion{Ca}{II} H\&K, \ion{Ca}{I}$\lambda$4226\AA, and H$\alpha$ lines obtained 
with the X-shooter spectrograph and the predicted line positions as a function 
of the magnetic field strength. Interestingly, the observed $\pi$ components 
for both \ion{Ca}{ii} lines appear weaker than the $\sigma$ components 
showing the effect of inclination of the field with respect to the 
line-of-sight. No significant variation in the line positions or intensities 
were noted between the three usable exposures.

\begin{figure*}[t!]
\includegraphics[width=1.00\columnwidth]{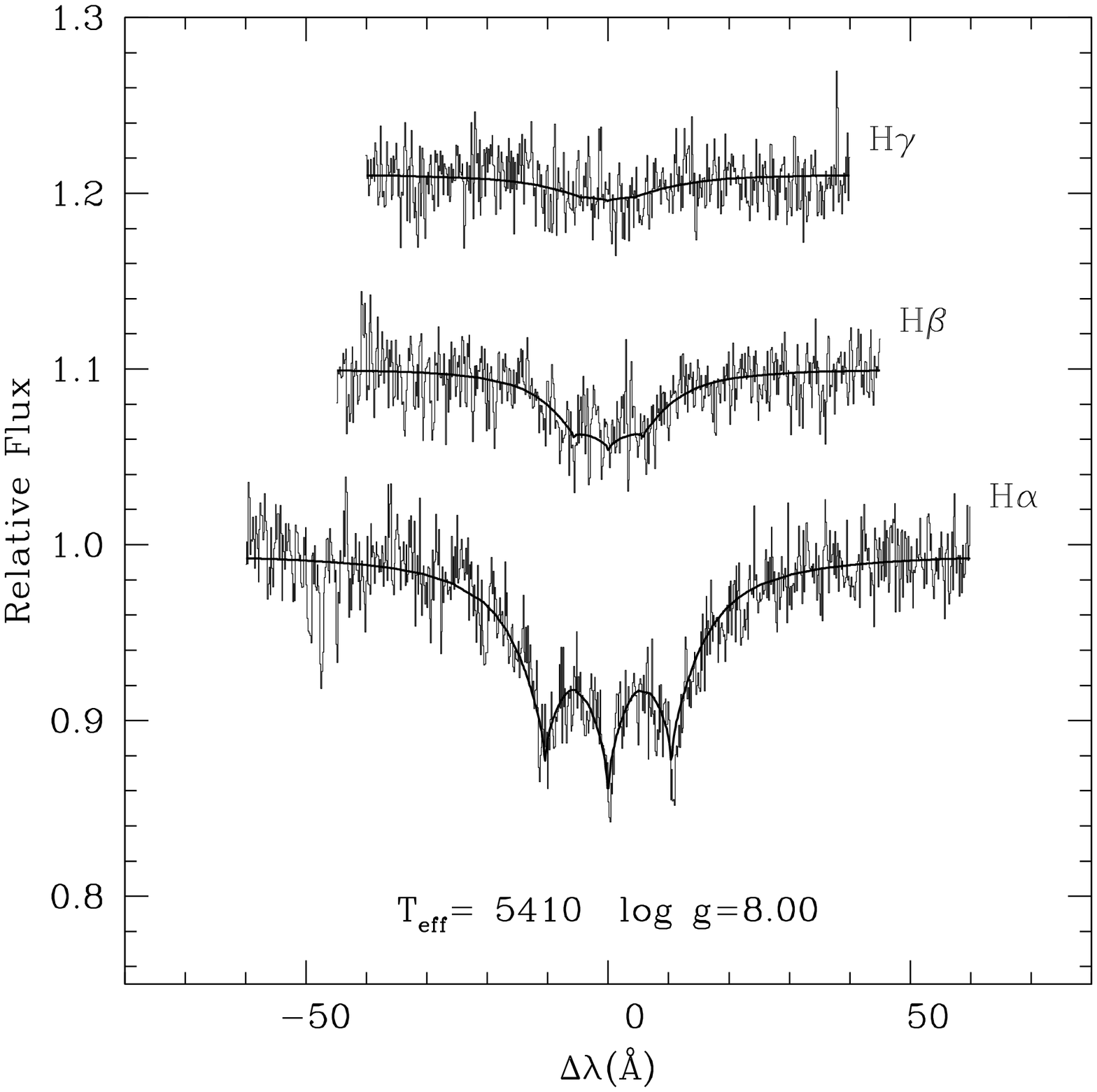}
\includegraphics[width=1.00\columnwidth]{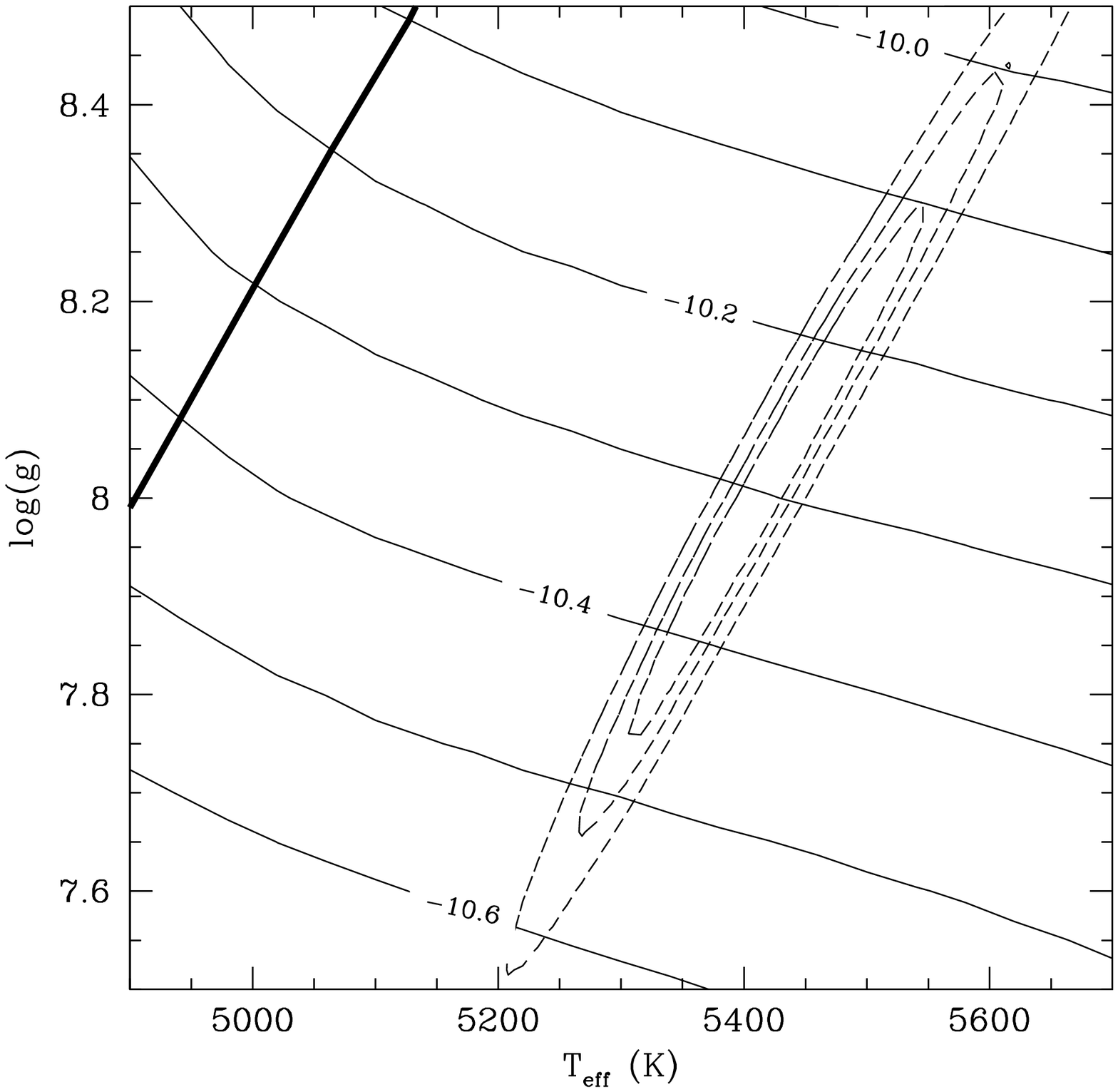}
\caption{Balmer lines (H$\alpha$ to H$\gamma$) and best-fitting model ($T_{\rm eff}=5410$ K,
$\log{g}=8.0$) showing prominent Zeeman-split H$\alpha$ line (left panel), and (right panel)
$\chi^2$ contours at 66, 90, and 99\% confidence (dashed lines) in the ($T_{\rm eff}$, $\log{g}$) plane. The measured calcium
abundance from \ion{Ca}{ii} lines ($\log{n{\rm (Ca)}/n{\rm (H)}}$) is drawn (full lines) as a function of the assumed temperature and surface gravity
and labelled with the logarithm of the abundance. The thick line shows the range of temperature and surface gravity where the abundance
measurements based on \ion{Ca}{i} or \ion{Ca}{ii} are consistent.
\label{fig-chi2}}
\end{figure*}

We fitted the ratio of the \ion{Ca}{II} $\pi$ and $\sigma$ line equivalent 
widths varying the inclination and determined $\psi = 60\pm3^\circ$. However,
including magneto-optical effects would likely decrease the measured inclination.

\subsection{Magnetic field and atmospheric parameters}

We determined the averaged surface magnetic field of NLTT~10480 using 
H$\alpha$ and the calcium lines. We first measured the centroids of the Zeeman 
components of \ion{Ca}{II}, \ion{Ca}{I} and H$\alpha$, where the spectrum was 
already adjusted to the solar system barycentre. We then fitted these lines to 
the predicted line positions by varying the magnetic field strength 
(assumed constant over the surface) and the 
velocity and by minimizing the $\chi^2$. For calcium, we determined a 
surface-averaged magnetic field of $B_s = 0.513\pm0.005$~MG with a velocity of 
$\varv = 50.8\pm3.0$ km~s$^{-1}$. Using H$\alpha$, we obtained 
$B_s = 0.525\pm0.005$~MG and $\varv = 51.8\pm3.0$ km~s$^{-1}$. The measurements 
are consistent within 2$\sigma$ and taking the average we found 
$B_s = 0.519\pm0.004$ and $\varv = 51.3\pm2$ km~s$^{-1}$. Subtracting the 
gravitational redshift ($\varv_{gr} = 29.3^{+34.2}_{-4.8}$ km~s$^{-1}$)
of the white dwarf from the measured velocity results in the actual line of 
sight velocity of the star $\varv_r = 22^{+34}_{-5}$ km~s$^{-1}$.

In principle, the effective temperature and surface gravity may be constrained
simultaneously by fitting the Balmer line profiles. The strong H$\alpha$ and 
weaker $\beta$ show the effect of Zeeman line splitting, while H$\gamma$ is 
extremely weak. Figure~\ref{fig-chi2} shows the best-fitting model to the 
Balmer lines that includes an approximate treatment of line opacities in a weak 
magnetic field. The 99\% confidence contour corresponds to an uncertainty of 
200~K in the temperature, and 0.5 dex in the logarithm of the surface gravity. 
The errors are statistical only and do not take into account possible 
systematic errors in the modelling of the line profiles.

\subsubsection{Abundance of heavy elements}

We determined the calcium abundance by fitting the \ion{Ca}{ii} line profiles 
using $\chi^2$ minimization techniques where we varied the calcium abundance at
each grid point in the ($T_{\rm eff}, \log{g}$) plane while fixing the magnetic field
strength at 0.519 MG (Sect. 3.3). Figure~\ref{fig-chi2} shows the resulting
abundance map: the calcium abundance
and its uncertainty depend on the adopted atmospheric parameters $T_{\rm eff}$ 
and $\log{g}$.

\begin{figure}
\includegraphics[width=1.00\columnwidth]{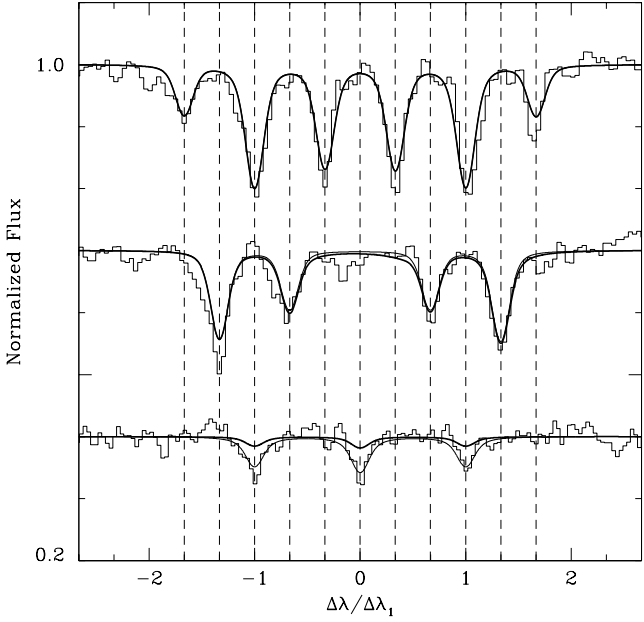}
\caption{X-shooter spectrum of the \ion{Ca}{i} (bottom) and \ion{Ca}{ii} 
(H \& K: middle and top, respectively) lines and 
best-fitting models at ($\log{g}=8$) $T_{\rm eff} = 5\,400$ K and 
$\log{n{\rm (Ca)}/n{\rm (H)}}=-10.3$ (thick lines), and $T_{\rm eff} = 4\,900$ K 
and $\log{n{\rm (Ca)}/n{\rm (H)}}=-10.45$ (thin lines). The wavelength scale
is normalized to $\Delta\lambda_1 \equiv 4.67\times10^{-7}\lambda^2B$, where 
$B=0.513$~MG. Therefore, the line positions 
match the factors $g_lm_l-g_um_u$ listed in Table~\ref{tbl-split} (vertical 
dashed lines).
All models assume $\psi=60^\circ$.
\label{fig-calcium}}
\end{figure}

Figure~\ref{fig-calcium} shows best-fitting models to the calcium lines
at $T_{\rm eff}=5\,400$ ($\log{g}=8$) and at a lower temperature of
4\,900 K ($\log{g}=8$). The calcium abundance varies between 
$\log{n{\rm (Ca)}/n{\rm (H)}}=-10.3$ (highest temperature) and $-10.45$ 
(lowest temperature), but the calcium ionization balance favours a lower
temperature than estimated with the Balmer lines (Fig.~\ref{fig-chi2}).
The discrepancy is partially resolved by increasing the metallicity of the
atmosphere. Using a set of heavy elements with low first-ionization potentials
and higher solar abundances including C, N, and O, then Na to Si, and, 
finally, K to Cu, we found that a high metallicity of $0.03\times$ solar
would restore the calcium ionization balance with an abundance of 
$\log{n{\rm (Ca)}/n{\rm (H)}}=-10.2$. However, a much lower abundance of heavy 
elements is present in the atmosphere of NLTT~10480.

\begin{figure}
\includegraphics[width=1.00\columnwidth]{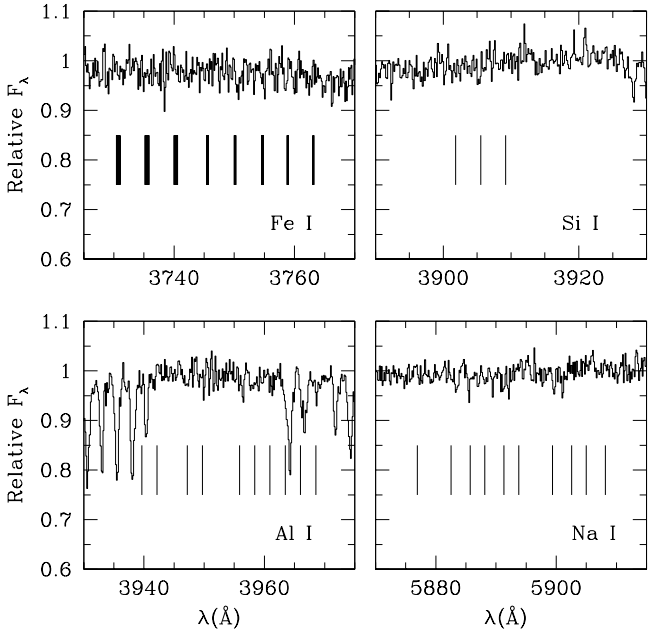}
\caption{Spectra in the vicinity of the
strongest \ion{Fe}{I} (3734.8638, 3749.4851, and 3758.2330\AA), \ion{Si}{I} (3905.523 \AA), \ion{Al}{I}
(3944.0060 and 3961.5200 \AA), and \ion{Na}{I} (5889.951 and 5895.924 \AA) lines
compared to the predicted Zeeman-split line positions
assuming a magnetic field of 0.519~MG.
\label{fig-other}}
\end{figure}

Figure~\ref{fig-other} shows the predicted location of the
strongest lines of \ion{Fe}{I}, \ion{Si}{I}, \ion{Al}{I} and \ion{Na}{I} in the
X-shooter spectrum. We calculated the position of the Zeeman-split 
lines for these elements assuming a magnetic field of 0.519~MG. For aluminium,
some weak lines appear to match the predicted positions. However, the
putative \ion{Al}{i}$\lambda$3967.8532 component should be accompanied by
stronger $\lambda$3957.7353 and $\lambda$3960.2458 components that are not 
clearly identified. Another possible identification for this feature is an 
interstellar Ca~H line, although Ca~K is not detected in the spectrum.
Spectra with higher signal-to-noise ratios and better resolution are needed to 
clarify this identification.

We estimated sodium, aluminium, silicon and iron abundance upper limits using 
the spectral ranges covered in Figure~\ref{fig-other}. We found
$\log{n{\rm (Na)}/n{\rm (H)}},\,\log{n{\rm (Al)}/n{\rm (H)}}$, and $\log{n{\rm (Fe)}/n{\rm (H)}} \la -9.3$,
while $\log{n{\rm (Si)}/n{\rm (H)}}\la -8.7$. The abundances relative to solar range from $2\times10^{-5}$
to $3\times10^{-4}$ times solar, or a few orders of magnitude below the level required
to significantly increase the electron density and alter the calcium ionization balance.

\section{Summary and discussion}

We found that the high proper-motion star NLTT~10480 is a rare example of cool 
white dwarfs with trace heavy elements and a weak magnetic field revealed in 
both the H$\alpha$ circular polarization spectrum and Zeeman line-splitting. 
Other examples of this class of white dwarfs are the DZ white dwarfs LHS~2534 
\citep{rei2001} and G~165-7 \citep{duf2006}, the DAZ white dwarfs G~77-50 
\citep{far2011} and possibly LTT~8381 \citep{koe2009b}. Based on 
independent diagnostics (Table~\ref{tbl-prop}), we estimated a temperature of 
$T_{\rm eff}=5\,200\pm200$ K and a surface gravity close to $\log{g}=8$.
However, we noted systematic differences in temperature measurements based on 
the calcium ionization ratio, the colour index, and the Balmer line profiles 
amounting to $\sim$400~K. The weaker \ion{Ca}{i} lines favour a lower 
temperature than estimated using Balmer lines alone. The temperature measured 
with the $V-J$ colour index also favours a lower temperature. We found that 
increasing the heavy-element contribution to the electron density helps restore 
the calcium ionization balance, but we also found that the required abundance 
exceeds upper limits on the abundance of Na, Al, Si, and Fe by a few orders of 
magnitude. We are left with the possibility that subtle effects on line 
formation (broadening parameters, magnetic-optical, ...) caused by the magnetic 
field may influence temperature measurements based on Balmer line profiles.

Although our modelling of the hydrogen line profiles takes into account
the effect of inclination, we neglected the magneto-optical effects and only
approximated the full solution of the radiative transfer equations that ought 
to include all Stokes parameters \citep[see][]{mar1981}. The effect of this 
approximation on the determination of the stellar parameters using Balmer line 
profile fitting may well amount to a few hundred degrees.

Fortunately, because ionized calcium is the dominant species, the abundance of 
calcium based on \ion{Ca}{ii} lines is not sensitive to temperature. On the 
other hand, as we have demonstrated, it does show a mild dependence on surface 
gravity because a higher electron pressure favours neutral calcium.

\begin{table}
\caption{Properties of NLTT~10480.\label{tbl-prop}}
\centering
\begin{tabular}{llc}
\hline\hline
Parameter & Measurement & Note \\
\hline
$B_S$         & $0.519\pm0.004$~MG & 1 \\
$\varv_r$     & $22^{+34}_{-5}$ km~s$^{-1}$ & 1 \\
$T_{\rm eff}$ & $5050\pm150$ K & 2 \\
              & $5400\pm200$ K & 3 \\
$\log{g}$     & $8.0\pm0.5$              & 3 \\
Abundance     & $\log{n{\rm (Na)}/n{\rm (H)}}\la -9.3$ & 4 \\
              & $\log{n{\rm (Al)}/n{\rm (H)}}\la -9.3$ & 4 \\
              & $\log{n{\rm (Si)}/n{\rm (H)}}\la -8.7$ & 4 \\
              & $\log{n{\rm (Ca)}/n{\rm (H)}} = -10.3\pm0.3$ & 4 \\
              & $\log{n{\rm (Fe)}/n{\rm (H)}}\la -9.3$ & 4 \\
$d$           & $33^{+9}_{-14}$ pc & 5 \\
  &  & \\
$U,V,W$       & $46^{+16}_{-31}, -66^{+29}_{-31}, -3^{+7}_{-33}$ km~s$^{-1}$ & 6 \\
\hline
\end{tabular}
\tablefoot{(1) Sect. 3.2; (2) $V-J$, Sect. 2; (3) Balmer lines, Sect. 3.3; (4) Sect. 3.3.1; (5) Based on apparent
and absolute $V$ magnitudes (Sect. 4); (6) Based on proper-motion and the estimated distance.}
\end{table}

Adopting conservative error bars for the temperature and surface gravity we 
calculated an absolute $V$ magnitude $M_V=14.9^{+0.9}_{-0.6}$ using the 
mass-radius relations of \citet{ben1999}. The distance modulus implies a
photometric distance $d=33^{+9}_{-14}$ pc. The object is relatively old with a 
cooling age $t_{\rm cool}=3.5-8.0$ Gyr, but with an uncertain mass 
($0.35-0.91\,M_\odot$). We determined the Galactic velocity vector $UVW$ 
(Table~\ref{tbl-prop}) using our distance estimate and radial velocity 
measurement (Sect. 3.3), and published proper-motion. We employed the algorithm 
of \citet{joh1987}. The kinematics imply membership to the old thin disk 
\citep{sio1988} consistent with the upper range of our age estimate.

The acquisition of broadband $UBV$ and $JHK$ photometry and of a parallax 
measurement should help determine the atmospheric parameters more precisely. 
The stellar radius, hence surface gravity measurement would be improved with a 
parallax measurement. Accordingly, the error on the calcium abundance 
measurement would be reduced. Accurate $JHK$ photometry would also allow us to 
investigate possible infrared excess and the presence of a debris disc.

The DAZ white dwarfs G~77$-$50 and G~174$-$74 are part of a survey including 
the coolest known DAZ white dwarfs \citep{zuc2003}. In cool convective white 
dwarfs, heavy elements diffuse below the mixed convective layers, and their 
presence in white dwarf atmospheres is transitory. \citet{koe2006} estimated 
the diffusion time-scale for various heavy elements. The accretion rate 
required to sustain a given mass fraction $X$ in the atmosphere is given by
\begin{equation}
\dot{M}_{\rm acc} = \frac{X}{X_{\rm acc}} \frac{M_{\rm cvz}}{\tau},
\end{equation} 
where $M_{\rm cvz}$ is the mass of the convection zone and $\tau$ the diffusion 
time-scale at the bottom of the convection zone, where diffusion is allowed to 
take place. The ratio $X/X_{\rm acc}$ is the ratio of the measured 
mass-fraction to the accreted mass fraction. Therefore, the mass accretion rate 
of any particular element is 
\begin{equation}
X_{\rm acc}\,\dot{M}_{\rm acc} = X\, \frac{M_{\rm cvz}}{\tau}.
\end{equation} 
Adopting, in the appropriate temperature range, a value for the slow-varying 
ratio $M_{\rm cvz}$ to diffusion time scale of 
$\log{(M_{\rm cvz}/\tau_{\rm Ca})}\approx -11.5$ in units of 
$M_\odot$\,yr$^{-1}$ or $=14.3$ in units of g\,s$^{-1}$ \citep{koe2006}, 
we estimated the mass accretion rate of calcium (in g\,s$^{-1}$) to be
\begin{equation}
\log{(X_{\rm acc,Ca}\,\dot{M}_{\rm acc})} = \log{(M_{\rm cvz}/\tau_{\rm Ca})} + \log{X_{\rm Ca}} = 5.60,
\end{equation}
where $X_{\rm Ca} \approx [A_{\rm Ca}\,n({\rm Ca})]/[A_{\rm H}\,n({\rm H})]= 2\times10^{-9}$,
and $A_{\rm Ca}$ and $A_{\rm H}$ are the atomic weights.
Assuming calcium is accreted as part of a solar-composition flow, the total
mass accretion rate (including hydrogen) is $5\times10^9$ g\,s$^{-1}$ or 
$8\times10^{-17}\ M_\odot$\,yr$^{-1}$.

Our measured error on the calcium abundance of $\pm0.3$ dex translates into
a similar error on the calculated accretion rate onto the white dwarf
surface. The true error may well be much larger. \citet{koe2009} 
considers that the application of the mixing-length theory to the structure of
convection zones may underestimate the mass of the mixed layers by orders of 
magnitude. The effect of ``under''-shooting below the convection zone may 
affect diffusion time-scale estimates. Therefore, the precision claimed in 
measuring abundance of parent bodies may be over-estimated. 

\citet{far2011} propose that the model of \citet{pot2010}\footnote{See also 
\citet{nor2011}} for the presence of
a magnetic field in post-common envelope (CE) binaries could also be applied
to CE episodes with planetary rather than stellar secondary components. Whether 
the magnetic field is acquired during such a process, or whether it is a fossil 
field cannot be ascertained for individual objects but rather from population 
studies \citep[see, e.g.,][]{kaw2004a,kaw2007,wic2005}.
In the case of NLTT~10480, which is old
($>3.5$ Gyr) with relatively short diffusion time-scales,
the present-day metallicity is not linked to the CE event that potentially 
generated the magnetic field, but more likely to a recent accretion event.

A low incidence of planetary systems would imply a low incidence of weak 
magnetic fields ($B\la 1$~MG). Current data indicate a low incidence of weak 
magnetic fields, and \citet{kaw2007} found that 6 out of 53 local white dwarfs 
($d \le 20$ pc) observed with sufficient accuracy to unveil fields weaker 
than 1~MG were found to harbour such a low field. Moreover, \citet{kaw2004a} 
found that low-field white dwarfs lack progenitors, a gap that could be filled 
with the CE-mechanism.

Early results from the {\it Kepler} survey also indicate a low incidence of 
very large planets in short-period orbits ($P\la 50$ days, or $a\la 0.2$ AU)
that are likely to trigger the field-generating CE events postulated by 
\citet{far2011} based on the model of \citet{pot2010}. Planets with sizes 
ranging from 8 to 32 $R_\oplus$ ($=0.7-2.9\,R_J$) may occur within 0.25~AU of 
solar-type stars with a frequency of 1.3\% \citep{how2011}. \citet{bor2011} 
quote a similar fraction for Jupiter-sized or larger planets within 0.2~AU 
based on the first data set from {\it Kepler}. Both studies noted a declining 
occurrence with increasing separation. Overall, large planets that are likely 
to participate in a CE phase may surround a few percents of white dwarf 
progenitors and generate, as observed, a similar percentage of low-field white 
dwarfs.

\begin{acknowledgements}

S.V. and A.K. are supported by GA AV grant numbers IAA300030908 and IAA301630901, respectively, and by GA \v{C}R grant number P209/10/0967.
A.K. also acknowledges support from the Centre for Theoretical
Astrophysics (LC06014). We thank the anonymous referee for 
several comments that improved the paper. This research has made use of the VizieR catalogue
access tool, CDS, Strasbourg, France.
This publication makes use of data products from the Two Micron All Sky Survey,
which is a joint project of the University of Massachusetts and the Infrared
Processing and Analysis Center/California Institute of Technology, funded by
the National Aeronautics and Space Administration and the National Science
Foundation.

\end{acknowledgements}

\end{document}